\documentclass[preprint, sd, showpacs, nofootinbib, superscriptaddress, longbibliography]{revtex4-1}
\usepackage{amsmath, amssymb, graphicx, subfigure, CJK} 
\usepackage{dcolumn, bm, multirow}
\usepackage{color, hyperref}
\usepackage[all]{hypcap}
\newcommand{\figurewidth}{0.8\textwidth}

\newcommand{\braket}[2]{\ensuremath{\langle#1\vert#2\rangle}}

\newcommand{\dspace}{\phantom{$\pm$0.00}}

\begin{document}
\begin{CJK*}{UTF8}{}

\title{Efficient electronic structure calculation for molecular ionization dynamics at high x-ray intensity}

\author{Yajiang Hao \CJKfamily{gbsn}(郝亚江)}
\email{haoyj@ustb.edu.cn}
\affiliation{Center for Free-Electron Laser Science, DESY, Notkestrasse 85, 22607 Hamburg, Germany}
\affiliation{Department of Physics, University of Science and Technology Beijing, Beijing 100083, P.~R.~China}
\affiliation{The Hamburg Centre for Ultrafast Imaging, Luruper Chaussee 149, 22761 Hamburg, Germany}

\author{Ludger Inhester}
\email{ludger.inhester@cfel.de}
\affiliation{Center for Free-Electron Laser Science, DESY, Notkestrasse 85, 22607 Hamburg, Germany}
\affiliation{The Hamburg Centre for Ultrafast Imaging, Luruper Chaussee 149, 22761 Hamburg, Germany}

\author{Kota Hanasaki}
\email{kota.hanasaki@cfel.de}
\affiliation{Center for Free-Electron Laser Science, DESY, Notkestrasse 85, 22607 Hamburg, Germany}
\affiliation{The Hamburg Centre for Ultrafast Imaging, Luruper Chaussee 149, 22761 Hamburg, Germany}

\author{Sang-Kil Son \CJKfamily{mj}(손상길)}
\email{sangkil.son@cfel.de}
\affiliation{Center for Free-Electron Laser Science, DESY, Notkestrasse 85, 22607 Hamburg, Germany}
\affiliation{The Hamburg Centre for Ultrafast Imaging, Luruper Chaussee 149, 22761 Hamburg, Germany}

\author{Robin Santra}
\email{robin.santra@cfel.de}
\affiliation{Center for Free-Electron Laser Science, DESY, Notkestrasse 85, 22607 Hamburg, Germany}
\affiliation{The Hamburg Centre for Ultrafast Imaging, Luruper Chaussee 149, 22761 Hamburg, Germany}
\affiliation{Department of Physics, University of Hamburg, Jungiusstrasse 9, 20355 Hamburg, Germany}

\date{\today}

\begin{abstract}
We present the implementation of an electronic-structure approach dedicated to ionization dynamics of molecules interacting with x-ray free-electron laser (XFEL) pulses. 
In our scheme, molecular orbitals for molecular core-hole states are represented by linear combination of numerical atomic orbitals that are solutions of corresponding atomic core-hole states. 
We demonstrate that our scheme efficiently calculates all possible multiple-hole configurations of molecules formed during XFEL pulses. 
The present method is suitable to investigate x-ray multiphoton multiple ionization dynamics and accompanying nuclear dynamics, providing essential information on the chemical dynamics relevant for high-intensity x-ray imaging.
%
%
\end{abstract}

\pacs{31.15.A-,31.15.ae,87.15.A-,02.70.-c}

\maketitle
\end{CJK*}

\section{Introduction}

The advent of x-ray free-electron lasers (XFELs)~\cite{McNeil10,Pellegrini12} opens up a new era in science and technology~\cite{Marangos11,Bucksbaum11,Ullrich12}.
Unprecedentedly ultraintense and ultrafast hard x-ray pulses generated from XFELs enable us to measure molecular structures on the atomic scale and to explore the structural dynamics on the femtosecond scale.
One of the most prominent XFEL applications is imaging of biological macromolecules.
X-ray crystallography with XFELs, after demonstration of the proof-of-principle~\cite{Chapman06}, has started to reveal previously unknown protein structure~\cite{Redecke13}, promising a breakthrough in structural biology (see reviews in Refs.~\cite{Fromme11,Schlichting12,Neutze12,Patterson14}).
Recent advances in time-resolved serial femtosecond crystallography~\cite{Aquila12,Kern13,Kern14,Kupitz14,Tenboer14} enable us to take a step forward towards molecular movies.
To investigate molecular structure and structural dynamics with XFELs, one needs to understand radiation damage dynamics---x-ray multiphoton ionization dynamics and accompanying nuclear dynamics.
Furthermore, the phase problem~\cite{Taylor03} is the bottleneck to reconstruct molecular structures in serial femtosecond crystallography as much as in conventional x-ray crystallography. 
To overcome the phase problem for x-ray crystallography with XFELs, one uses conventional phasing technique at intermediate x-ray intensity~\cite{Barends14}, or one takes an advantage of the large degree of ionization at high x-ray intensity.
The latter brings in high-intensity phasing (HIP) methods~\cite{Galli14a}, including high-intensity multiwavelength anomalous diffraction~\cite{Son11e,Son13} and high-intensity radiation damage induced phasing~\cite{Galli15}.
The HIP techniques require detailed description of deep-inner-shell decay dynamics of heavy atoms influenced by the molecular environment.
Therefore, understanding of radiation damage dynamics is the key for successful molecular imaging.

Modeling of biological macromolecules exposed to XFEL radiation was pioneered by the seminal work of Neutze \textit{et al.}~\cite{Neutze00}.
Since then, there have been several computational tools to address molecular imaging problems, for example, classical molecular dynamics model~\cite{Jurek04,Hau-Riege13}, particle-in-cell approach~\cite{Hau-Riege12,Varin12}, transport model~\cite{Hau-Riege04,Ziaja06a}, Coulomb complex model~\cite{Gnodtke11,Gnodtke12}.
Some of these methods have been recently applied to ion fragment spectra~\cite{Murphy14} and electron spectra~\cite{Camacho-Garibay14} from C$_{60}$ molecules interacting with intense x-ray pulses.
So far, most computational methods have been based on the independent-atom model or the plasma model.
The description of the molecular environment is omitted by assumption or incorporated in an \textit{ad hoc} manner.
When a molecule absorbs x-ray photons, inner-shell multiple ionization induces fragmentation dynamics~\cite{Carlson66a,Dunford12a}.
Chemical bonds are weakened and electrons and holes rearrange before the molecule breaks apart~\cite{Erk13,Schnorr14,Erk14}.
Detailed electronic structure calculations for chemical bonding and charge rearrangement are thus crucial to describe molecular effects in modeling of the XFEL--matter interaction.

The electronic response of atoms and molecules to XFEL pulses is in essence characterized by multiphoton multiple ionization dynamics~\cite{Young10,Hoener10,Santra14}. 
The \textsc{xatom} toolkit~\cite{Son11a,Son12f} has been developed to simulate the XFEL--atom dynamical interaction and the ionization dynamics model has been tested with a series of experiments~\cite{Doumy11,Rudek12,Fukuzawa13,Rudek13,Motomura13}.
The unprecedentedly large number of x-ray photons delivered by an XFEL pulse induces sequential x-ray absorptions, creating a variety of different $q$-hole configurations for each charge state $+q$.
To simulate ionization dynamics, one needs to calculate photoionization cross section, Auger rate, and fluorescence rate for each configuration and solve a set of coupled rate equations for the time-dependent populations of the configurations~\cite{Rohringer07,Young10,Makris09}.
The $q$-hole electronic configurations are energetically highly excited, and theoretical treatment of such highly-excited states is challenging.  
Another challenge is the complexity of the ionization dynamics.  
Even for the atomic case, one must solve more than 20 million coupled rate equations for Xe $L$-shell-initiated ionization dynamics~\cite{Fukuzawa13}.
To address this formidable problem, a Monte-Carlo approach has been proposed for solving the rate equations~\cite{Son12f,Rudek12} and sampling the most probable configurations~\cite{Fukuzawa13}.  
In this scheme, the electronic structure is calculated for every single configuration selected on the fly as part of the Monte Carlo sampling.  
For the molecular case, the complexity increases even further because of the degrees of freedom associated with atomic motions, so the Monte Carlo approach seems to be the only way to simulate molecular response to an intense XFEL pulse.  
However, the most important question still remains: how to describe the electronic structure of molecules on the fly for the Monte Carlo method?

Here we present a new efficient method for electronic structure calculation of polyatomic molecules and implement a dedicated toolkit, \textsc{xmolecule}.
The proposed method is well suited for calculations of molecular multiple-hole configurations that are formed during x-ray multiphoton ionization dynamics.
To efficiently describe molecular orbitals of core-hole configurations, the method employs atomic orbitals as basis functions that are numerical solutions of atomic core-hole states, calculated by \textsc{xatom}~\cite{Son11a}.  
For any given molecular electronic configuration and any given molecular geometry, 
\textsc{xmolecule} calculates molecular orbitals and orbital energies, which are essential components for dynamical simulations of x-ray multiphoton multiple ionization. 
We demonstrate that \textsc{xmolecule} is capable to calculate the whole spectrum of multiple-hole configurations at a given molecular geometry and potential energy surfaces for given multiple-hole configurations of molecules. 
Also performance scalability with the system size is discussed. 
In this paper, we focus on the implementation of a molecular electronic-structure approach.
Calculating cross sections and rates and solving coupled rate equations to simulate ionization dynamics will be described elsewhere.
Having achieved these results, \textsc{xmolecule} aims to play a key role in molecular imaging at high x-ray intensity.   

The paper is organized as follows. 
Section~\ref{sec:method} formulates our scheme to calculate molecular multiple-hole configurations.
It includes theoretical and computational schemes for basis function generation with numerical atomic orbitals, multicenter integration on a molecular grid, and direct Coulomb integral evaluation.
In Sec.~\ref{sec:results} we show benchmark calculations for \textsc{xmolecule}, and then numerical results for the potential energy curves of various electronic configurations of carbon monoxide, and single- and double-core ionization potentials of several polyatomic molecules.
We discuss the scalability of our scheme to a molecular size of hundreds of atoms. 
This is followed by the conclusion in Sec.~\ref{sec:conclusion}.

\section{Computational methods}\label{sec:method}

\subsection{The Hartree-Fock-Slater method}

We consider a molecular system composed of $N_\text{atom}$ atoms with $N_\text{elec}$ electrons. 
The $A$th nuclear charge and coordinates are denoted by $Z_{A}$ and ${\mathbf{R}_{A}}$, respectively.
The molecular charge state $+q$ is given by $q = \sum_A Z_A - N_\text{elec}$.
We use the Hartree-Fock-Slater (HFS) method in which molecular orbitals (MO), $\psi_i(\mathbf{r})$, and orbital energies, $\varepsilon_i$, are obtained by solving the effective single-electron Schr\"odinger equation (atomic units are used unless specified otherwise),
\begin{equation}
\left[ -\frac{1}{2} \nabla^{2} + V_{\mathrm{ext}}(\mathbf{r})+V_{H}(\mathbf{r})+V_{X}(\mathbf{r}) \right] \psi_i(\mathbf{r}) = \varepsilon_i \psi_i(\mathbf{r}),
\label{HFSEqs}
\end{equation}
where $V_{\mathrm{ext}}(\mathbf{r})$ is the external potential due to the nuclei,
\begin{equation}
V_{\mathrm{ext}}(\mathbf{r}) = - \sum_{A}\frac{Z_{A}}{\left|\mathbf{r}-\mathbf{R}_{A}\right|}, 
\end{equation}
and the Hartree potential $V_{H}(\mathbf{r})$ represents the classical Coulomb interaction among the electrons,
\begin{equation}\label{V_H}
V_{H}(\mathbf{r})= \int d^{3}r' \, \frac{\rho(\mathbf{r}')}{\left|\mathbf{r}-\mathbf{r}'\right|},
\end{equation}
and the last term $V_{X}(\mathbf{r})$ represents the exchange interaction, which is approximated by the Slater exchange potential~\cite{Slater51}, 
\begin{equation}\label{V_X}
V_{X}(\mathbf{r})=-\frac{3}{2} \left[ \frac{3}{\pi}\rho(\mathbf{r}) \right]^{\frac{1}{3}}.
\end{equation}
The electronic density $\rho(\mathbf{r})$ is obtained by the sum of squared MO's weighted by the occupation numbers~$\{ n_{i}\}$ as
\begin{equation}
\rho(\mathbf{{r}})=\sum_{i}n_{i} \left| \psi_{i}(\mathbf{{r}}) \right|^2,
\label{rho}
\end{equation}
where $n_{i}\in\{0,1,2\}$.
In contrast to conventional ground-state electronic structure calculations, in which the $N_\text{elec}$ spin-orbitals with the lowest energies are filled, we consider all possible $\lbrace n_{i} \rbrace$ subject to $\sum_{i}n_{i} = N_\text{elec}$, in order to take account of electronic excited states representing $q$-hole configurations.

The total energy within the HFS method is given by the sum of the nucleus--nucleus repulsion energy and the electronic energy,
\begin{align}\label{total_E}
E_\text{total} =
& \sum_{A<B} \frac{Z_A Z_B}{|\mathbf{R}_A - \mathbf{R}_B|}
+ \sum_i n_i \varepsilon_i 
 - \frac{1}{2} \int \! \! d^3 r \! \int \! \! d^3 r' \, \frac{\rho(\mathbf{r}) \rho(\mathbf{r}')}{ | \mathbf{r} - \mathbf{r}' | }
+ \frac{3}{8} \! \left( \frac{3}{\pi} \right)^\frac{1}{3} \!\!\! \int d^3 r\, \rho(\mathbf{r})^\frac{4}{3}.
\end{align}

\subsection{Linear combination of numerical atomic orbitals}\label{Subsec:LCAO}
For atomic systems, the orbital is represented with spherical harmonics as
\begin{equation}\label{AO}
\phi_{nlm}(\mathbf{r})=\frac{u_{nl}(r)}{r}Y_{lm}(\theta,\varphi),
\end{equation}
where $n$, $l$, and $m$ are the principal quantum number, the orbital angular momentum quantum number, and the associated projection quantum number, respectively.
The radial wavefunction $u_{nl}(r)$ can be solved by a numerical grid-based method. 
The \textsc{xatom} toolkit~\cite{Son11a} has been developed to solve the atomic HFS equation.
By employing the generalized pseudospectral (GPS) method~\cite{Yao93a,Tong97a} and imposing a spherically symmetric potential, \textsc{xatom} accurately calculates $u_{nl}(r)$ for a given $(n,l)$-subshell, and accordingly $\phi_\mu(\mathbf{r})$ for a given $\mu \equiv (n,l,m)$.
This numerical atomic orbital has been used to successfully calculate multiple-hole configuration formed during x-ray multiphoton ionization dynamics in the atomic case~\cite{Son11a,Santra14}.

For molecular systems, we employ the linear combination of atomic orbitals (LCAO) scheme to construct molecular orbitals,
\begin{equation}
\psi_{i}(\mathbf{r}) = \sum_{\mu} C_{\mu i} \phi_{\mu}(\mathbf{r}),
\label{LCAO}
\end{equation}
where $\phi_{\mu}(\mathbf{r})$ is the $\mu$th atomic orbital (AO) and $C_{\mu i}$ is the coefficient of the $\mu$th AO for the $i$th MO. 
Using Eq.~\eqref{LCAO} transforms the self consistent field (SCF) Eq.~\eqref{HFSEqs} into the corresponding Roothaan-Hall equation~\cite{Roothaan51}, 
\begin{equation}
\mathbf{HC}=\mathbf{SCE},
\label{SCFEqs}
\end{equation}
where $\mathbf{E}$ is a diagonal matrix of MO energies and $\mathbf{C}$ is the MO coefficient matrix.
The elements of the Hamiltonian matrix $\mathbf{H}$ and the overlap matrix $\mathbf{S}$ are given as
\begin{eqnarray}
\label{Hmatrix}
H_{\mu\nu} & = & \int d^{3}r \, \phi_{\mu}(\mathbf{r}) \left[ -\frac{1}{2} \nabla^{2} + V_{\mathrm{eff}}(\mathbf{r}) \right]\phi_{\nu}(\mathbf{r}),
\\
\label{Smatrix}
S_{\mu\nu} & = & \int d^{3}r \, \phi_{\mu}(\mathbf{r}) \phi_{\nu}(\mathbf{r}),
\end{eqnarray}
where the effective potential $V_{\mathrm{eff}}(\mathbf{r})\equiv V_{\mathrm{ext}}(\mathbf{r})+ V_{H}(\mathbf{r}) + V_{X}(\mathbf{r})$.
Equation~\eqref{SCFEqs} is solved in a self-consistent manner.
To accelerate convergency, we employ the direct inversion in the iterative subspace (DIIS) method~\cite{Pulay80,Pulay82}. 
When we encounter convergence problems at large bond distances, where the energy gap between the highest occupied valence orbital (HOMO) and the lowest unoccupied virtual orbital (LUMO) is very small, we apply level shifts~\cite{Saunders73} in the SCF iterations.

\begin{figure*}[]
\includegraphics[width=\textwidth]{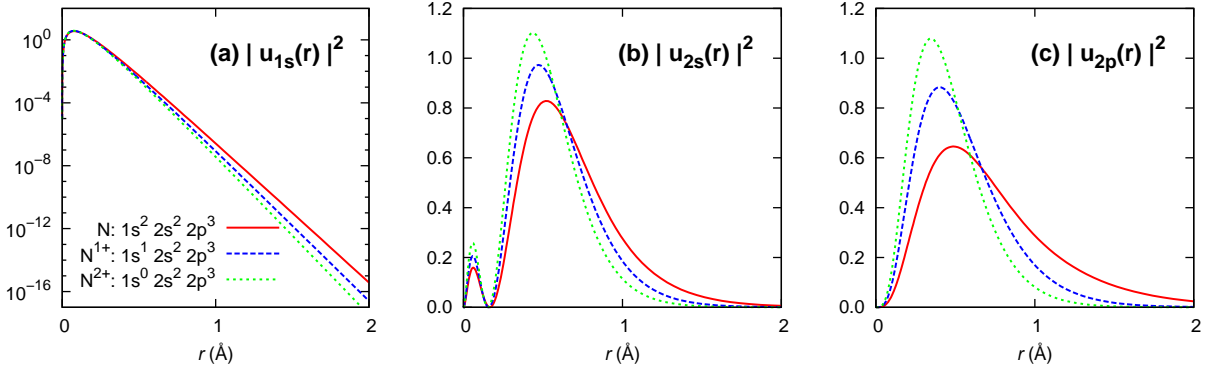}
\caption{Numerical atomic orbitals for different core-hole states of the nitrogen atom.}\label{NAO_N}
\end{figure*}

Here our choice of basis set for the LCAO scheme is the numerical atomic orbitals (NAO) obtained by \textsc{xatom} described above.
In Fig.~\ref{NAO_N}, we plot the squared radial function $|u_{nl}(r)|^2$ for the $1s$, $2s$, and $2p$ orbitals of the ground state of the neutral nitrogen (N) atom, the single-core-hole (SCH) state of N$^{+}$, and the double-core-hole (DCH) state of N$^{2+}$, respectively. 
Comparison among different core-hole states shows significant deformation of valence orbitals in states with core holes. 
To cover these effects efficiently in the molecular calculation, we use NAOs that are numerical solutions of the corresponding atomic core-hole states.
For instance, N$_2^{2+}$ with one core hole at each atomic site (a DCH state) is calculated with basis functions optimized for N$^+$($1s^{-1}$) on both N atoms, whereas N$_2^{2+}$ with a single-site DCH state is calculated with basis functions optimized for N$^{2+}$($1s^{-2}$) on which the core hole is located and basis functions optimized for neutral N on the other side.
In this way, we expect core-hole MOs are well described by core-hole-adapted NAOs.

To achieve utmost efficiency towards complex ionization dynamics, we employ the \emph{minimal} basis set.
Each AO with $(n,l,m)$ in Eq.~\eqref{AO} corresponds to a single basis function.
Fully or partially occupied $(n,l)$-subshells contribute to a set of basis functions and each $l$ gives $(2l+1)$ basis functions ($|m| \leq l$).
For example, the N atom has $1s$, $2s$, and $2p$ (partially) occupied subshells, which constitute 5 basis functions ($\phi_{1s}$, $\phi_{2s}$, $\phi_{2p_x}$, $\phi_{2p_y}$, and $\phi_{2p_z}$) in total.
This basis set is denoted as [2s1p].
According to the minimal-basis-set scheme, the chemical elements from B to Ne have the same number of basis functions ($N_\text{basis}$=5).
In Section~\ref{sec:benchmark}, we will discuss limitations and extensions of the minimal-basis-set scheme.

\subsection{Molecular grid and multicenter integration}

Equations~\eqref{Hmatrix} and \eqref{Smatrix} require evaluation of the corresponding integrals in three dimensions.
In our case, the $\phi_\mu({\bf r})$ and $\phi_\nu({\bf r})$ are represented with a radial grid and spherical harmonics.
To perform 3D integrals involving many atomic centers, we employ the multicenter integration proposed by Becke~\cite{Becke88}.
Molecular grid points are constructed as a combination of sets of atomic grid points.
Each set of atomic grid points, centered at one of the nuclei, consists of $N_r$ radial grid points and $N_\text{ang}$ angular grid points.
The radial grid points are exactly the same as those used for NAO calculations with the GPS method~\cite{Yao93a,Tong97a}.
The angular grid points are obtained by the Lebedev grid scheme~\cite{Lebedev76} with an angular momentum cutoff at $l_\text{max}$.
The number of angular grid points is approximately given by $N_\text{ang} \approx 4 (l_\text{max}+1)^2 / 3$.
A detailed description of constructing multicenter molecular grid points is found in Refs.~\cite{Son09b,Son11}.
We use an atomic radial grid size ($r_\text{max}$) large enough ($\sim$10~\AA) so that the atomic grids of many neighboring atoms overlap with each other.
In principle, different atomic grid parameters can be used for individual atoms in a molecule.
For convenience, however, we use the same grid parameters for all atoms.
Then the total number of molecular grid points is given by $N_\text{grid} = N_\text{atom} \times N_r \times N_\text{ang}$.

Becke's multicenter integration scheme~\cite{Becke88} introduces a set of smooth nuclear weight functions $\{ w_A(\mathbf{r}) \}$, subject to the constraint $\sum_{A} w_A(\mathbf{r})=1$. 
The nuclear weight functions are generated by the third-order polynomial cutoff profile in the \emph{fuzzy cell} scheme~\cite{Becke88}.
Then any integral of a given function $f$ can be evaluated by the sum of individual atomic integrals,
\begin{align}
I & = \int d^3 r \, f(\mathbf{r}) = \sum_A \int d^3 r \, f(\mathbf{r}) w_A(\mathbf{r})
\approx \sum_A \int\limits_A d^3 r_A \, f(\mathbf{r}_A) w_A(\mathbf{r}_A),
\end{align}
where $\mathbf{r}_A \equiv \mathbf{r} - \mathbf{R}_A$.
Each atomic integral can be readily performed using the spherical coordinate system of $\mathbf{r}_A$, centered at the $A$th atom,
\begin{equation}
I_A = \int\limits_A d^3 r_A \, f(\mathbf{r}_A) w_A(\mathbf{r}_A) \approx \sum_{k \in A} f(\mathbf{r}_k) w_A(\mathbf{r}_k) w_k,
\end{equation}
where $k$ is the index of the grid points of the $A$th atom and $w_k$ is defined as a product of the radial Legendre-Gauss-Lobatto quadrature weights~\cite{Canuto88,Boyd01} and the angular Lebedev quadrature weights~\cite{Lebedev76}.

\subsection{Implementation of direct Coulomb integrals}\label{Subsec:Implementation_of_direct_Coulomb}

In electronic structure calculations, one of the most time-consuming parts is the evaluation of electron repulsion integrals. 
In order to achieve fast calculation within a desired accuracy, we have developed a multipole expansion scheme with an adaptive cut off. 
First the integral involved in the Hartree potential in Eq.~\eqref{V_H} can be decomposed into individual atomic integrals,
\begin{equation}
V_{H}(\mathbf{r}) 
= \int d^{3}r' \, \frac{\rho(\mathbf{r}')}{\left|\mathbf{r}-\mathbf{r}'\right|}
= \sum_A \int\limits_A d^{3}r'_A \, \frac{\rho_{A}(\mathbf{r}'_A)}{\left|\mathbf{r}_A-\mathbf{r}'_A\right|},
\end{equation}
where $\rho_{A}(\mathbf{r}) \equiv \rho(\mathbf{r}) w_{A}(\mathbf{r})$.
Each single-center density $\rho_{A}(\mathbf{r})$ can then be regarded as the atomic contribution to the total electronic density.
To implement the integral we expand the single-center density with real spherical harmonics $Y_{lm}(\theta,\varphi)$ as
\begin{equation}\label{eqn:VlmDecomposition}
\rho_{A}(\mathbf{r}_A) = \sum_{l=0}^{l_\text{max}} \sum_{m=-l}^l \rho_{lm}^A(r_A) Y_{lm}(\theta_A,\varphi_A),
\end{equation}
where $\rho_{lm}^A(r)$ is the $(l,m)$-component of the spherical expansion,
\begin{equation}\label{eqn:rhoAexpansion}
\rho_{lm}^A(r_A) = \int_0^{2\pi} \! d\varphi_A \int_0^\pi \! d\theta_A \, \sin\theta_A \, \rho_{A}(\mathbf{r}_A) Y_{lm}(\theta_A,\varphi_A).
\end{equation}
With this single-center decomposition and spherical harmonic expansion of the electronic density, $\rho_{lm}^A(r)$,
the Hartree potential in Eq.~\eqref{V_H} is obtained as
\begin{equation}
V_{H}(\mathbf{r}) = \sum_A \sum_{l,m} V_{lm}^A(r_A) Y_{lm}(\theta_A,\varphi_A).
\end{equation}
where $V_{lm}^A$ is given by
\begin{align}\label{eqn:radialintegrals}
V_{lm}^A(r_A) = \frac{4\pi}{2l+1} \int_{0}^{r_\text{max}} \! dr'_A \, {r'_A}^{2} \frac{{r_<}^{l}}{r_>^{l+1}} \rho_{lm}^A(r'_A),
\end{align} 
where $r_< = \min( r'_A, r_A )$ and $r_> = \max( r'_A, r_A )$.
This radial integral is numerically evaluated in combination with various truncation methods (see the Appendix).

\subsection{Molecular electronic configuration}\label{SubSec:Molecular_electronic_configuration}

Keeping the energetically lowest orbitals doubly occupied, the SCF procedure obtains the HFS solution for the electronic ground state.
In order to obtain a solution for an excited electronic state of a $q$-hole configuration, each molecular orbital has to be assigned a specific occupation number.
This can be done, as in the ground state calculation, by identifying the orbitals by their HFS energy eigenvalue.
However, during the SCF iterations the energetic order of MOs may change.
Thus, identifying the orbitals by ordering them according to their HFS energy eigenvalue may lead to failure of the above SCF procedure or yield a solution for a different electronic state than required. 
This is called variational collapse~\cite{Bagus65,Jensen87,Gilbert08}.

To prevent this situation, we employ a variant of the maximum overlap method~\cite{Gilbert08}.
In the maximum overlap method, the desired excited electronic state is specified by a set of initial guess orbitals $\{ \psi_j^\text{guess} \}$ in combination with a set of occupation numbers $\{ n_j \}$.
In each SCF iteration, the occupation number $n_i$ of the calculated orbital $\psi_i$ is chosen according to 
its projection onto the subspace spanned by the guess orbitals with respective occupation number.
Specifically, we calculate the overlap of the $i$th current MO with the $j$th guess MO, 
\begin{equation}
O_{ij} = \braket{ \psi_i }{ \psi_j^\text{guess} } 
= \sum_{\mu,\nu} C_{\mu i} \tilde{S}_{\mu \nu} C_{\nu j}^{\mathrm{guess}},
\end{equation}
where $\tilde{S}_{\mu \nu} = \int \! d^3 r \, \phi_\mu(\mathbf{r}) \phi_\nu^\text{guess}(\mathbf{r})$.
Note that the basis set for the initial guess orbitals is not necessarily the same as the one used for the expansion of the actual molecular orbitals, because different NAOs can be used for different $q$-hole configurations. 
Therefore, $\tilde{S}_{\mu \nu} $ can be different from the overlap matrix $S_{\mu \nu}$ defined in Eq.~\eqref{Smatrix}.
Then, the projections of the $i$th orbital into the span of the guess orbitals for the unoccupied ($n$=0), singly occupied ($n$=1), and doubly occupied ($n$=2) cases are given by
\begin{equation}\label{MOM_P}
P_i^{(n)} =  \sum_{j} \left| O_{ij} \right|^2,
\end{equation}
where $j$ runs over all initial guess orbitals whose occupation number $n_j$ equals $n$.
To preserve the character of the required electronic configuration during the SCF procedure, we choose the set of the occupation numbers of the current orbitals, $\{n_i\}$, such that $\sum_i P_i^{(n_i)}$ is maximized, while the total number of doubly and singly occupied orbitals is maintained.

This procedure to determine the orbital occupation critically depends on the initial guess MOs. 
Thus, it is essential that the provided guess MOs $\{\psi_j\}$ together with the provided occupation numbers $\{n_j\}$ describe a wavefunction that is close to the required solution.
For the calculations performed here, we choose initial guess MOs obtained from a previous calculation for a lower ionized electronic state or for the same electronic state with an altered molecule geometry. 
For the single-core-hole state in N$_2$, we obtain a localized core hole on a specific nucleus by performing a Boys-orbital-localization procedure~\cite{Kleier74} of the two guess core orbitals.
Having obtained a converged solution, we verify that the obtained set of MOs is indeed close to the initial guess, by inspecting the individual overlap $O_{ij}$.

\section{Results and discussion}\label{sec:results}

\subsection{Benchmark calculations}\label{sec:benchmark}

\begin{figure}
\centering
\includegraphics[width=\figurewidth]{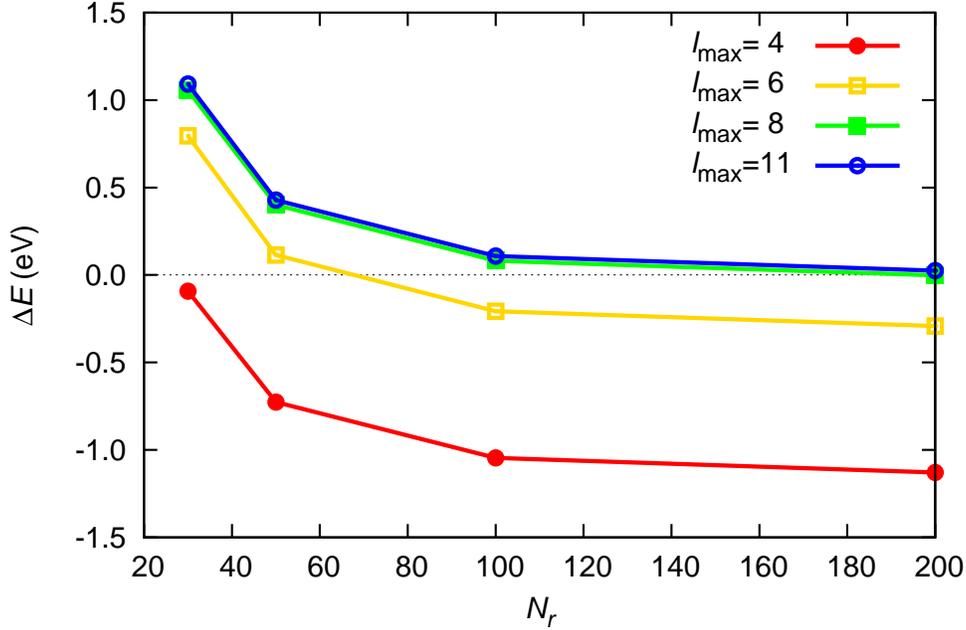}\label{fig:H2O_convergence}
\caption{Convergence of the total HF energy with respect to the number of grid points.
$N_r$ is the number of radial grid points and $l_\text{max}$ controls the number of angular grid points per atom.  
The ground-state energy calculation of H$_{2}$O with RHF/6-31G is performed using the numerical multicenter integration, and $\Delta E$ is the difference from the result obtained using the analytic Gaussian integrals.}
\end{figure}

We first estimate the accuracy of our calculations using the numerical multicenter integration in comparison with conventional calculations using the analytic Gaussian integration by \textsc{gamess}~\cite{GAMESS}.
Here we employ the 6-31G Gaussian basis set~\cite{Hehre72} to calculate the SCF-level ground-state energy of a water molecule. 
The internuclear distance of $R(\mathrm{OH})=0.957$~\AA\ and the bond angle of $\angle(\mathrm{HOH})=104.48^\circ$ are used.
Only in this test we employ the restricted Hartree-Fock (RHF) method instead of the HFS method, in order to directly compare with the \textsc{gamess} results.
Figure~\ref{fig:H2O_convergence} shows that our numerical calculations converge to the \textsc{gamess} results as the number of radial grid points per atom ($N_r$) and the number of angular grid points per atom (determined by $l_\text{max}$) are increased.
The total number of molecular grid points for $N_r$=50 and $l_\text{max}$=8 is $3\times50\times110 = 16500$.
The maximum radius $r_\text{max}$=20~a.u.\ and the GPS mapping parameter~\cite{Yao93a,Tong97a} $L$=1~a.u.\ are used.
Note that all grid parameters utilized here provide a numerical accuracy $| \Delta E | < 1.5$~eV.
If chemical accuracy is required (typically 1~kcal/mol $\approx$ 0.04~eV), our study for the water molecule shows that it is achievable with $N_r \geq 200$ and $l_\text{max} \geq 11$, keeping the same $L$ and $r_\text{max}$.
As to be shown in Sec.~\ref{sec:PEC}, the energy scale of x-ray-induced dynamics of highly-charged molecules will extend into the keV regime.
Therefore, the worst grid parameters (for example, $N_r$=30 and $l_\text{max}$=4) shown in Fig.~\ref{fig:H2O_convergence} would be sufficient to describe the molecular ionization dynamics at high x-ray intensity.

\begin{figure}
\centering
\subfigure[\ neutral N$_2$]{\includegraphics[width=3.2in]{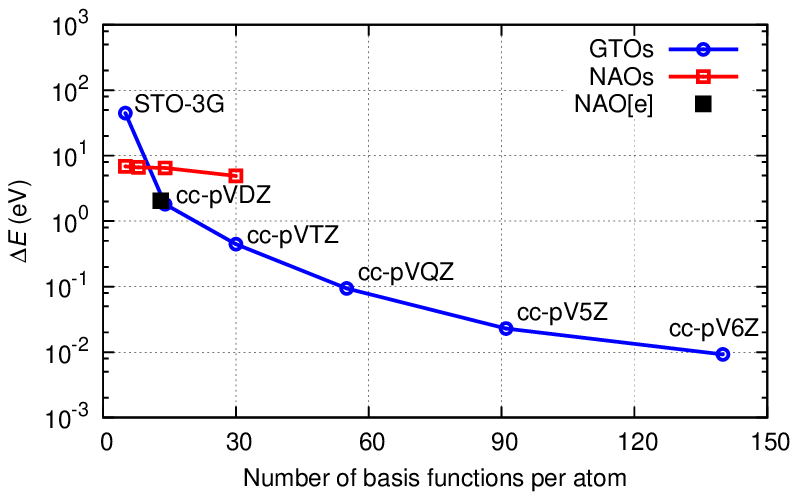}}
\subfigure[\ quadruple-core-hole N$_2^{4+}$]{\includegraphics[width=3.2in]{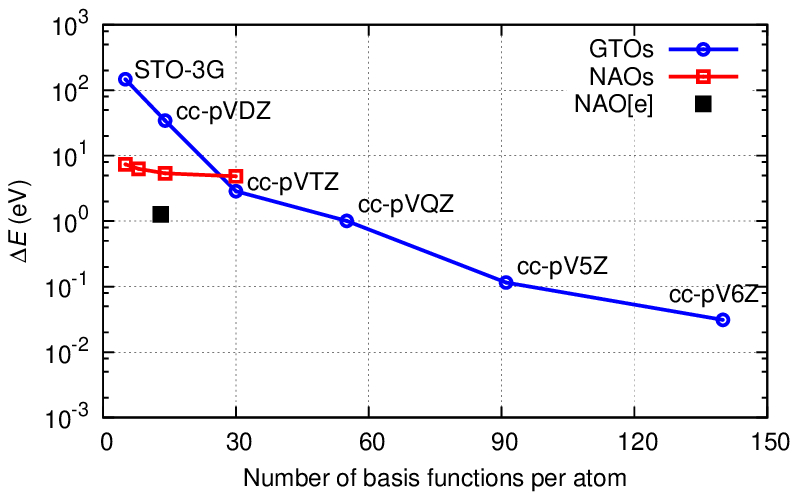}} \label{fig:NAO_performance}
\caption{Comparison of convergency in total energy with respect to the number of basis functions, using the GTO scheme and the NAO scheme: (a) neutral N$_2$ and (b) QCH N$_2^{4+}$.
$\Delta E$ is defined by the total energy difference from the complete basis set limit (see text).}
\end{figure}

We next examine the performance of our NAO basis set scheme. 
In Fig.~\ref{fig:NAO_performance}, we show the calculated HFS energy of (a) the ground state of neutral N$_2$ molecule with NAOs optimized for neutral N atom and (b) the quadruple-core-hole (QCH) state of N$_{2}^{4+}$ ion with NAOs optimized for the DCH state of N$^{2+}$. 
The internuclear distance $R$=1.096~\AA\ is fixed.
$N_r = 200$, $L=1$~a.u., $r_\text{max}=20$~a.u., and $l_\text{max} = 11$ are used.
The results are shown together with those obtained by the equivalent calculations using conventional Gaussian-type-orbital (GTO) basis sets of different sizes (STO-3G~\cite{Hehre69} and a series of Dunning's correlation-consistent basis sets~\cite{Dunning89}; All GTO basis sets are obtained from the EMSL Basis Set Library~\cite{Feller96}).
$\Delta E$ is the energy difference from the total energy calculated with the uncontracted version of cc-pV6Z, [16s10p5d4f3g2h1i] with 161 basis functions, which is considered here the complete basis set limit.
Thus $\Delta E$ indicates the numerical error due to lack of basis functions.

In both Figs.~\ref{fig:NAO_performance}(a) and (b), one can see that the minimal NAO basis set is superior to the conventional minimal basis set of STO-3G, illustrating that fully optimized NAOs are a practical choice for the basis set in the LCAO scheme.
Also Fig.~\ref{fig:NAO_performance} shows convergency of GTOs with respect to the number of basis functions.
Interestingly, the conventional GTOs for QCH N$_2^{4+}$ perform almost one order of magnitude less accurate than GTOs for neutral N$_2$.
The reason is that GTOs are optimized to be used for neutral ground-state calculations.
In contrast, NAOs optimized for corresponding atomic $q$-hole configuration provide similar accuracy for both neutral N$_2$ and QCH N$_2^{4+}$.
Thus NAO functions provide an ideal basis set for our minimal-basis-set HFS scheme.

To improve accuracy, we try to increase the number of NAOs in a systematic manner by including unoccupied atomic orbitals with higher $(n,l)$ such as $3s$, $3p$, and so on.
As shown in Fig.~\ref{fig:NAO_performance}, the NAOs are somewhat inefficient to achieve higher accuracy by simply extending to higher $(n,l)$, as previously reported in Ref.~\cite{Blum09}.
This is attributed to the fact that additional series of higher $(n,l)$-orbitals, whose mean square radius is far from the atomic center, are inefficient for representing bonding molecular orbitals. 
Instead, we propose a scheme for adding compact $p$-type and $d$-type functions to the minimal NAO basis set in order to improve the description of chemical bonding.
Additional functions are constructed by use of radial wavefunctions of occupied subshells multiplied by $r$, where $r$ is the radial coordinate in the atomic system.
For the chemical elements from B to Ne, the $p$-type functions are $u_{2s}(r) Y_{1m}(\theta,\varphi)$, where $m = 0, \pm1$, and the $d$-type functions are $u_{2p}(r) Y_{2m}(\theta,\varphi)$, where $m = 0, \pm 1, \pm 2$.
By adding these functions, as denoted by extended NAO (NAO[e]) and as marked with the black rectangle in Fig.~\ref{fig:NAO_performance}, the accuracy is much improved; the total energy of neutral N$_2$ is close to the cc-pVDZ level and the total energy of QCH N$_2^{4+}$ is close to the cc-pVQZ level.
The number of basis functions for NAO[e] is only 13 per atom, whereas cc-pVQZ has 55 basis functions.
There have been several approaches for extension of the minimal NAO basis set~\cite{Blum09,Anglada02}, where additional basis functions are constructed in a schematic way.

\subsection{Potential energy curves for various hole configurations}\label{sec:PEC}

\begin{figure}[tbp]
\includegraphics[width=\figurewidth]{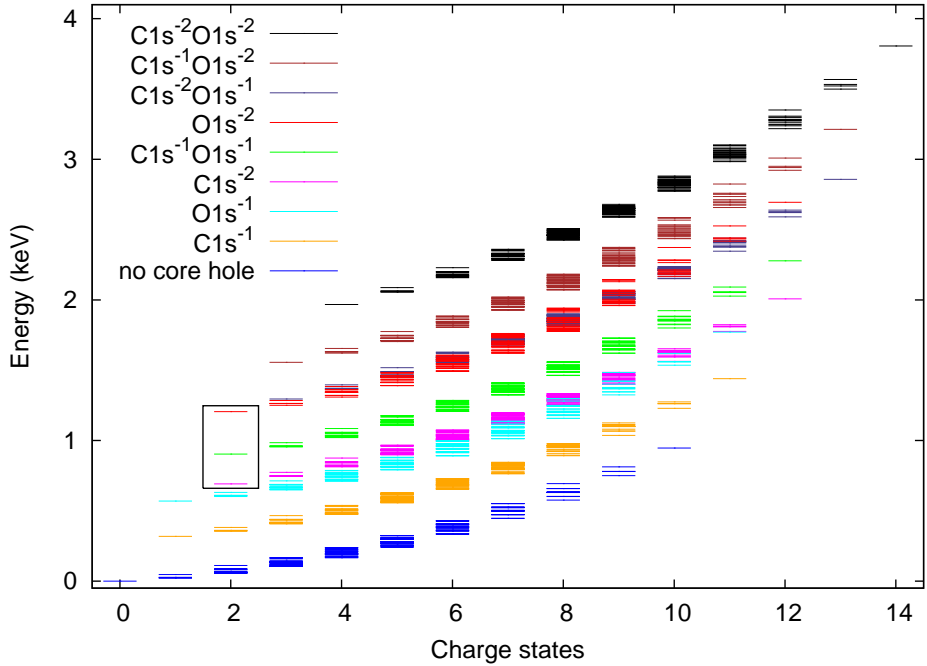}
\caption{Spectrum of total energies for various electronic states of CO, which are accessible by x-ray multiphoton ionization.
The colors indicate different core hole configurations.
Potential energy curves of the DCH states inside the box will be shown in Fig.~\ref{PEC_CO_DCH}.
\label{E_Q_CO}}
\end{figure}

Figure~\ref{E_Q_CO} shows the HFS total energies in Eq.~\eqref{total_E} using core-hole-adapted NAO basis functions for all possible $q$-hole configurations that can be accessed by x-ray multiphoton ionization of the neutral carbon monoxide molecule.
The internuclear distance $R$=1.128~\AA\ is fixed, and the grid parameters of $N_r$=50, $L$=1~a.u., $r_\text{max}$=20~a.u., and $l_\text{max}$=8 are used.
For convenience, the figure shows these configurations grouped into charge states. 
The lowest horizontal line for each charge state indicates the ground-state energy for a given charge $+q$.
This figure then illustrates how much energetically excited the $q$-hole configurations are.
For example, the energy of DCH CO$^{2+}$ (O$1s^{-2}$) is about 1~keV higher than the ground-state energy of CO$^{2+}$.
Ionization dynamics induced by intense x-ray pulses may occur step by step, visiting lots of these electronic states.
Therefore it is crucial to efficiently calculate this set of electronic states of $q$-hole configurations. 

\begin{figure*}[]
\includegraphics[width=\textwidth]{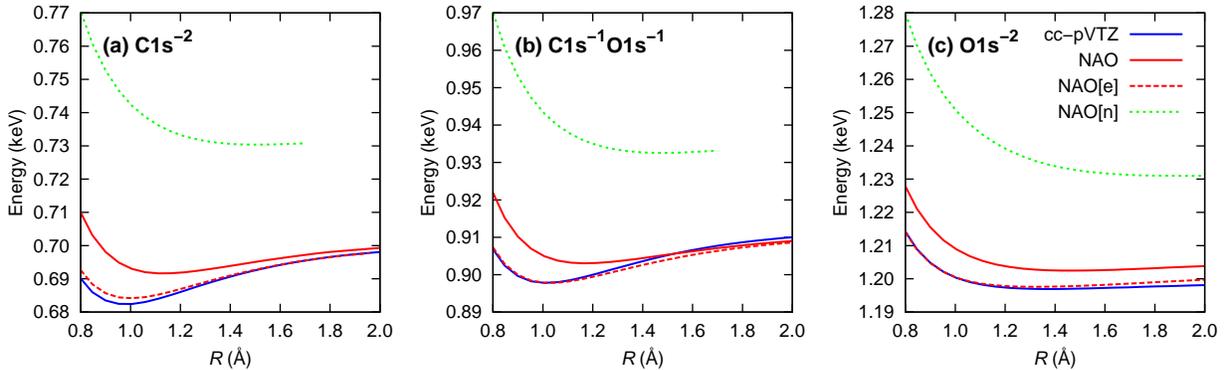}
\caption{Potential energy curves of CO$^{2+}$ double-core-hole states as a function of the internuclear distance $R$. 
Energy is given relative to the ground-state energy of neutral CO.
} \label{PEC_CO_DCH}
\end{figure*}

We further investigate the behavior of the potential energy curves (PEC) obtained using the NAO basis set. 
In Fig.~\ref{PEC_CO_DCH}, we show the calculated HFS total energies for three different types of CO$^{2+}$ DCH states: 
(a) C$1s^{-2}$, (b) C$1s^{-1}$O$1s^{-1}$, and (c) O$1s^{-2}$.
The solid red line indicates PECs calculated with core-hole-adapted NAO[e].
The dashed red line indicates PECs with core-hole-adapted NAO without additional functions.
Both results are compared with the solid blue line calculated with the conventional cc-pVTZ basis set.
Previous theoretical studies of core-hole states suggested that calculations of the cc-pVTZ level are reasonably converged~\cite{Tashiro10,Carravetta13}.
Our NAO[e] scheme reproduces well the cc-pVTZ results, even though the size of NAO[e] ($N_\text{basis}$=13) is much smaller than that of cc-pVTZ ($N_\text{basis}$=30).
For comparison, we also plot PECs with NAOs optimized for neutral ground-state atoms, denoted by NAO[n], which shows a trend similar to what a conventional STO-3G minimal basis set would be.
The NAO[n] results represent a poor estimate of PECs due to missing the core-hole effect on orbitals.
On the other hand, PECs from NAOs, which are optimized for atomic core-hole states, show dramatic improvement over NAO[n], even though NAO and NAO[n] have the same number of basis functions ($N_\text{basis}$=5).

\subsection{Single- and double-core ionization potentials of molecules}

To further test the accuracy of our calculation scheme we compare core ionization potentials for a series of small molecules obtained from the HFS calculation.  
The molecular geometries are taken from Ref.~\cite{CCCBDB} and the grid parameters are the same as those used in Sec.~\ref{sec:PEC}.
We derive the single-core ionization potential from the HFS orbital energy of a neutral ground-state calculation using NAOs with and without additional basis functions.
The double-core ionization potential is calculated as the sum of the first and the second core ionization potential,
where the second core ionization potential is taken from the orbital energy of the SCH state calculation.
For the DCH states with core holes on different nuclear sites, thus, two values are obtained for the two different 
ionization sequences. 

\begin{table*}[tbp]
\caption{Single core hole and double core hole ionization potentials in eV.
The molecular geometries are taken from Ref.~\cite{CCCBDB}.}
\begin{ruledtabular}
\begin{tabular}{llrrrr}
Molecule	& Configuration			& Present (NAO)		& Present (NAO[e]) 	& CASSCF~\cite{Tashiro10}	& Exp. \\  
\hline 
CO		& O$1s^{-1}$		 		& 537.43\dspace		&533.80\dspace		&542.82		& 542.5$^a$ \\ 
		& C$1s^{-1}$		 		& 295.81\dspace		&289.84\dspace		&296.36		& 296.5$^b$ \\ 
		& O$1s^{-2}$		 		& 1139.12\dspace	&1136.43\dspace		&1176.56	& 			\\           
		& C$1s^{-2}$		 		& 647.50\dspace		&636.89\dspace		&664.42		& 667.9$^b$ \\ 
		& C$1s^{-1}$O$1s^{-1}$		& 850.70$\pm$2.28	& 840.34$\pm$0.52	& 855.20	& 855.3$^b$ \\
\hline 
LiF 	& F$1s^{-1}$	     		& 663.77\dspace   	&670.23\dspace		&688.04		& 691.8$^c$ \\ 
		&Li$1s^{-1}$		 		& 59.34\dspace    	&58.56\dspace		&65.33		&  61.9$^d$ \\ 
		& F$1s^{-2}$				& 1403.81\dspace  	&1420.99\dspace		&1481.50 	& \\ 
		& Li$1s^{-2}$				& 154.84\dspace   	&153.19\dspace		&172.60 	& \\ 
		& Li$1s^{-1}$F$1s^{-1}$	 	& 735.13$\pm$2.72   &739.48$\pm$1.06	&763.28     & \\ 
\hline 
N$_{2}$	& N$1\sigma_\mathrm{g}^{-1}$ 	& 409.57\dspace		&403.30\dspace		&411.03	& 409.9$^e$ \\  
		& N$1\sigma_\mathrm{u}^{-1}$	& 409.54\dspace		&403.26\dspace		&410.93	&           \\  
		& N$1s^{-2}$		 			& 878.29\dspace		&868.83\dspace		&901.16	& 903.2$^f$ \\  
		& N$1s_A^{-1}$N$1s_B^{-1}$		& 836.96$\pm$0.02   &823.87$\pm$0.02	&836.44 &           \\ 
\hline 
N$_{2}$O	& O$1s^{-1}$		 	& 537.81\dspace		&534.72\dspace		&542.54		& 541.4$^g$ \\ 
		& N$_t1s^{-1}$		 		& 408.70\dspace		&403.66\dspace		&408.61		& 409.0$^f$ \\ 
		& N$_c1s^{-1}$		 		& 413.74\dspace		&407.89\dspace		&412.52		& 412.5$^e$ \\ 
		& O$1s^{-2}$		 		& 1138.96\dspace	&1136.23\dspace		&1173.25 	& \\ 
		& N$_t1s^{-2}$		 		& 874.42\dspace		&866.83\dspace		&893.93		& \\ 
		& N$_c1s^{-2}$		 		& 883.76\dspace		&875.54\dspace		&902.31 	& \\ 
		& O$1s^{-1}$N$_t1s^{-1}$	& 961.29$\pm$0.25	&951.47$\pm$0.25	&963.27		& \\ 
		& O$1s^{-1}$N$_c1s^{-1}$ 	& 964.30$\pm$0.40	&954.53$\pm$0.37	&965.62		& \\ 
		& N$_t1s^{-1}$N$_c1s^{-1}$	& 836.55$\pm$0.01	&825.25$\pm$0.11	&833.22		& 834.2$^f$\\ 
\end{tabular}
\label{IPs_list}
\end{ruledtabular}
$^a$Ref.~\cite{Puttner99}; 
$^b$Ref.~\cite{Berrah11};
$^c$Ref.~\cite{Hudson94};
$^d$Ref.~\cite{Haensel68};
$^e$Ref.~\cite{Alagia05};
$^f$Ref.~\cite{Salen12};
$^g$Ref.~\cite{Bakke80} \\
\end{table*}

Table~\ref{IPs_list} lists the ionization potentials compared with the values obtained from complete-active-space SCF (CASSCF) calculations~\cite{Tashiro10} and experimental values~\cite{Puttner99,Berrah11,Hudson94,Haensel68,Alagia05,Salen12}.
For our calculations of two-site DCH states, a mean value and a deviation are listed for two values from the different ionization sequences.
For the CASSCF results of two-site DCH states, only triplet spin states are listed and the difference between singlet and triplet states is smaller than 0.7~eV.
Note that N$_2$O is a linear molecule N$_t$---N$_c$---O, where N$_t$ indicates the terminal N atom and N$_c$ means N at the center.
As can be seen, the CASSCF results~\cite{Tashiro10} show agreement within less than $4$~eV with the available experimental values.
The single ionization potentials we extract from the much simpler HFS calculation using the minimal NAO basis set show for all molecules a similar agreement within $5.1$~eV, except F$1s^{-1}$ in LiF ($28$~eV).
For the DCH states, where the core holes are located on different nuclear sites, with the minimal NAO basis set 
we also see a similar agreement within $7$~eV to the CASSCF values and, where available, the experimental values.
Again, LiF is an exception showing a much larger discrepancy of $\simeq 30$~eV.
For the DCH state with core holes on the same nucleus we find a systematically larger disagreement of about $20$--$30$~eV (for F$1s^{-2}$ in LiF $78$~eV).

The inclusion of the $p$-type and $d$-type functions in the basis set leads in most cases to a larger deviation 
to the literature values than the results obtained with the minimal basis set.
For these calculations we get ionization potentials that tend to be lower than the literature values (from $3.4$~eV for Li$1s^{-1}$ to $60.5$~eV for F$1s^{-2}$ in LiF).
Clearly, the extended NAO basis set should improve the quality of the electronic structure model, as the electronic wave function has more flexibility.
Thus, we conclude that the good agreement with the minimal basis set might be an artifact due to cancellation of errors.

For the results obtained with the larger basis set, we attribute the remaining deviations to the literature values mainly to 
relaxation energy contributions associated with the core hole electron removal.
The applied scheme of taking orbital energies as ionization potentials cannot account for these effects.
For core holes on the same nuclear site, where the core hole relaxation contributions are particularly strong, we see the strongest deviations ($18.2$--$60.5$~eV).
Also, the extreme deviations for LiF may be explained from these contributions:
The core hole on the F atom in LiF shows a particular large core hole relaxation effect, whereas for the core hole on Li it is very small~\cite{Tashiro10}.

\subsection{Performance scaling}

Our implementation of \textsc{xmolecule} aims for large-scale molecular calculations, especially for a large number of repeated calculations where time and resources available for each calculation are severely limited.
At the same time, it requires the capability of calculating a moderate-size systems in order to describe molecular-environment effects.
Here we demonstrate the performance scalability of our scheme toward molecular calculations with a few hundred atoms.
Our grid-based method has the potential to achieve linear scaling in the number of atoms~\cite{Goedecker99,Junquera01,Watson04,Havu09}.

In the HFS method, the two-body interaction is divided into the exchange interaction and the direct Coulomb interaction.
The former is replaced with the local density approximation, and the latter is treated with the Hartree potential as described in Sec.~\ref{Subsec:Implementation_of_direct_Coulomb}.
The computational complexity of the Hartree potential is $O(N_\text{grid}^2)$, where $N_\text{grid}$ is linearly proportional to $N_\text{atom}$, because the potential $V_H(\mathbf{r})$ in Eq.~\eqref{V_H} contains the integral over molecular grid points and has to be evaluated at every single molecular grid point.
By introducing the truncation methods described in the Appendix, this complexity can be reduced to $O(N_\text{grid} N_\text{atom})$.
These truncation schemes do not change the quadratic scaling behavior with respect to $N_\text{atom}$, but reduce the actual computational time by several times (for example, a factor of two in our following calculations).

Another truncation can be made in the evaluation of one-body matrix elements in Eqs.~\eqref{Hmatrix} and \eqref{Smatrix}.
Both $H_{\mu \nu}$ and $S_{\mu \nu}$ are decomposed into atomic contributions by the multicenter integration: $H_{\mu \nu} \approx \sum_A H_{\mu \nu}^A$ and $S_{\mu \nu} \approx \sum_A S_{\mu \nu}^A$.
We define an AO pair $\phi_\mu(\mathbf{r}) \phi_\nu(\mathbf{r})$ and its contribution to each atomic grid,
\begin{equation}\label{truncation_S}
Q_{\mu \nu}^A = \int\limits_A d^3 r_A \left| \phi_\mu(\mathbf{r}_A) \phi_\nu(\mathbf{r}_A) \right| w_A(\mathbf{r}_A).
\end{equation}
Then we set $H_{\mu \nu}^A$ and $S_{\mu \nu}^A$ to zero if $Q_{\mu \nu}^A < \varepsilon$, where $\varepsilon$ is a control parameter.
The complexity of the integrals in Eqs.~\eqref{Hmatrix} and \eqref{Smatrix} is $O(N_\text{basis}^2 N_\text{grid})$, where both $N_\text{basis}$ and $N_\text{grid}$ are linearly proportional to $N_\text{atom}$.
By using our truncation scheme described above, we can reduce it to a quadratic behavior with respect to $N_\text{atom}$.

\begin{figure}
\includegraphics[width=\figurewidth]{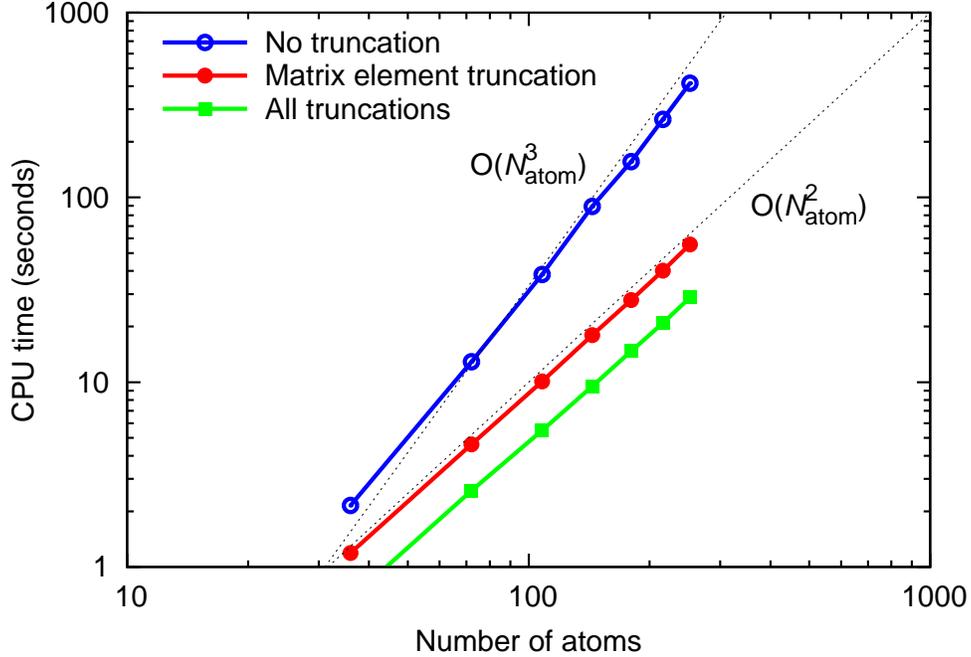}\label{fig:C24H12n_timePerformance}
\caption{Performance scaling with respect to the molecular size.
The $y$ axis is the CPU time per SCF iteration in seconds, and the $x$ axis is the number of atoms in stacked (C$_{24}$H$_{12}$)$_n$ molecules.
The dotted lines with $O(N_\text{atom}^2)$ and $O(N_\text{atom}^3)$ indicate a quadratic behavior and a cubic behavior, respectively, with respect to the number of atoms, $N_\text{atom}$.
}
\end{figure}

Figure~\ref{fig:C24H12n_timePerformance} shows the size dependence of the computation time of \textsc{xmolecule} with the current truncation schemes.
We calculate the HFS ground state of C$_{24}$H$_{12}$ molecule (coronene) in its equilibrium molecular geometry taken from Ref.~\cite{CCCBDB}.
And we perform calculations for $n$ such molecules ($n$=$1,\dots,7$) stacked in the vertical direction with an interlayer separation of 3.3~\AA.
The minimal NAO basis set is used with $N_r$=20, $L$=1~a.u., $r_\text{max}$=10~a.u., and $l_\text{max}$=4.
The $y$ axis is the CPU time per SCF iteration in seconds on a lab workstation (Intel Xeon X5660 2.80~GHz), and the $x$ axis indicates the number of atoms in the stacked (C$_{24}$H$_{12}$)$_n$ molecule.
When all truncations are off (blue curve), the computational performance shows close to a cubic dependence.
On the other hand, when the truncation method of Eq.~\eqref{truncation_S} is applied with $\varepsilon=10^{-3}$ (red curve), the scaling shows a quadratic dependence on the system size.
Note that when the truncation of Eq.~\eqref{truncation_S} is used, the complexity of the matrix element calculations is reduced to a quadratic relation, while the Hartree potential calculation becomes the most time-consuming step, which is also governed by a quadratic scaling.
The difference in the total energy between the calculations with and without this truncation is less than 0.14~eV/atom, whereas the truncated calculation is about 7.5 times faster than the calculation with no truncation.
The calculation with 216 atoms ($n$=6) takes 40 seconds per single SCF iteration on the lab workstation.
The whole computation time takes about 14 minutes including the overhead costs for numerical grid construction and 12 SCF iterations.
%
When additional truncation schemes for the Hartree potential (see the Appendix) are applied with $\varepsilon_0$=0.1 and $\varepsilon_1$=0.01 (green curve), the complexity is a bit reduced towards a linear relation and the errors in the total energy are less than 0.93~eV/atom.
The actual computational time per iteration is improved by a factor of two for the 216-atom case.

\section{Conclusion}\label{sec:conclusion}
In summary, we present a new method to calculate various multiple-hole electronic states for polyatomic molecules that may be formed by x-ray multiphoton ionization dynamics at high x-ray intensity.
The method is based on the Hartree-Fock-Slater method, employing the linear combination of atomic orbitals (LCAO) scheme, where numerical atomic orbitals (NAO) are used as a minimal basis set for molecular orbital calculations.
Usage of NAOs has two advantages over conventional Gaussian-type basis functions.
First, NAOs are obtained from numerical solutions for atomic core-hole states at the same computational level.
Second, accuracy and efficiency of numerical integration with NAOs are controllable by grid parameters and truncation schemes.
The NAOs presented here are accurately solved by using the numerical grid-based method that is implemented in the \textsc{xatom} toolkit.

Using core-hole-adapted NAOs, molecular orbitals for core-hole states are efficiently calculated.
We present benchmark calculations for multiple-core-hole states of N$_2$.
The NAO results show consistent accuracy for different charge states, which is not the case for conventional basis sets that are optimized for neutral systems.
We demonstrate that our scheme is able to calculate all possible configurations that may be formed by removing zero, one or more electrons from the ground-state configuration of neutral CO molecule.
The electronic state during x-ray multiphoton ionization dynamics may visit several of these multiple-hole configurations, which are energetically excited by about 4~keV with respect to the ground-state configuration of neutral CO.
For molecular and ionization dynamics during XFEL pulses, we need not only all different multiple-hole states but also potential energy surfaces for individual electronic states.
For double-core-hole states of CO$^{2+}$, we calculate potential energy curves with core-hole-adapted NAOs, in good agreement with converged results with respect to the basis-set size.
Also we present single- and double-core-hole ionization potentials for several molecules in comparison with available theoretical and experimental data.

Efficient electronic structure calculations for molecules are essential for dynamical modeling of molecules at high x-ray intensity.
We have implemented \textsc{xmolecule} to make a step toward dynamical simulation of molecular imaging with XFELs.
Calculations of photoionization cross sections, fluorescence rates, and Auger rates for all possible configurations formed during molecular ionization dynamics are in progress.

\begin{acknowledgments}
We thank Oriol Vendrell for helpful discussions.
This work has been supported by the excellence cluster `The Hamburg Centre for Ultrafast Imaging -- Structure, Dynamics and Control of Matter at the Atomic Scale' of the Deutsche Forschungsgemeinschaft.
Yajiang Hao is supported by the National Natural Science Foundation of China (Grant No.~11004007) and the Fundamental Research Funds for the Central Universities of China.
\end{acknowledgments}

\section*{Appendix}\label{appendix}
\appendix*
Here, we introduce truncation schemes on $V_{lm}^A(r_A)$.
The upper bound of $\left| V_{lm}^A \right|$ is given by
\begin{align}
\left| V_{lm}^A(r_A) \right|
&= \frac{4\pi}{2l+1}
\left|
\int_{0}^{r_A} \! dr'_A \, {r'_A}^{2} \frac{{r'_A}^{l}}{r_A^{l+1}} \rho_{lm}^A(r'_A) +
\int_{r_A}^{r_\text{max}} \! dr'_A \, {r'_A}^{2} \frac{r_A^{l}}{{r'_A}^{l+1}} \rho_{lm}^A(r'_A)
\right|,
\nonumber
\\
&\leq \frac{4\pi}{2l+1} \left[ 
\left| \frac{1}{r_A} \int_{0}^{r_A} \! dr'_A \, {r'_A}^{2} ( \frac{r'_A}{r_A} )^l \rho_{lm}^A(r'_A) \right| + 
\left| \frac{1}{r_A} \int_{r_A}^{r_\text{max}} \! dr'_A \, {r'_A}^{2} ( \frac{r_A}{r'_A} )^{l+1} \rho_{lm}^A(r'_A) \right|
\right]
\nonumber
\\
&\leq \frac{4\pi}{2l+1} \left[ 
\frac{1}{r_A} \int_{0}^{r_A} \! dr'_A \, {r'_A}^{2} \left| \rho_{lm}^A(r'_A) \right| + 
\frac{1}{r_A} \int_{r_A}^{r_\text{max}} \! dr'_A \, {r'_A}^{2} \left| \rho_{lm}^A(r'_A) \right|
\right]
\nonumber
\\
&= \frac{4\pi}{2l+1} \cdot \frac{1}{r_A} \int_0^{r_\text{max}} \! d r'_A \, {r'_A}^2 \left| \rho_{lm}^A(r'_A) \right|.
\end{align}
Then we define
\begin{equation}
d_{lm}^A = \int_0^{r_\text{max}} \! d r'_A \, {r'_A}^2 \left| \rho_{lm}^A(r'_A) \right|,
\end{equation}
to be used as a truncation indicator.
Note that the number of electrons in the $A$th atomic electronic density is given by $Q_A = \int \! d^3 r \, \rho_A(\mathbf{r}) = \sqrt{ 4 \pi } d_{00}^A$. 
Within the atom we consider higher multipole moments of the density to be less relevant.
Thus, if $d_{lm}^A$ is small enough in comparison with $d_{00}^A$, then the contribution of $l$ and $m$ is truncated, i.e.,
\begin{equation}
V_{lm}^A(r_A) \to 0 \quad \text{when }\frac{d_{lm}^A}{d_{00}^A} < \varepsilon_1,
\end{equation}
where $\varepsilon_1$ is a truncation control parameter.

Another truncation is that if the distance from the origin of the $A$th atom is large enough, the Hartree potential contributed from $A$ is approximately evaluated by the monopole only and all $l > 0$ contributions are truncated, i.e.,
\begin{equation}
V_{lm}^A(r_A) \to 0 \quad \text{when }r_A > r_c,
\end{equation}
where $r_c$ is a cut-off radius given by $r_c = Q_A / {\varepsilon_0} = {\sqrt{4 \pi} d_{00}^A} / \varepsilon_0$.
Here $\varepsilon_0$ is another truncation control parameter.


\end{document}